\documentclass[twocolumn,showpacs,prc]{revtex4}
\usepackage{graphicx}
\usepackage{dcolumn}
\usepackage{bm}
\begin{document}i
\title{Concerning the Nature of the Cosmic Ray Power Law Exponents}
\author{A. Widom and J. Swain}
\affiliation{Physics Department, Northeastern University, Boston MA USA}
\author{Y.N. Srivastava}
\affiliation{Physics Department, University of Perugia, Perugia IT}

\begin{abstract}
We have recently shown that the cosmic ray energy distributions as detected 
on earthbound, low flying balloon or high flying satellite detectors can be 
computed by employing the heats of evaporation of high energy 
particles from astrophysical sources. In this manner, the experimentally well 
known power law exponents of the cosmic ray energy distribution have been theoretically 
computed as 2.701178 for the case of ideal Bose statistics, 3.000000 for the 
case of ideal Boltzmann statistics and 3.151374 for the case of ideal Fermi statistics. 
By ``ideal'' we mean virtually zero mass (i.e. ultra-relativistic) and noninteracting.
These results are in excellent agreement with the experimental indices of 2.7 with 
a shift to 3.1 at the high energy  $\sim $ PeV  ``knee'' in the energy distribution.
Our purpose here is to discuss the nature of cosmic ray power law exponents obtained by 
employing conventional thermal quantum field theoretical models such as quantum 
chromodynamics to the cosmic ray sources in a thermodynamic scheme wherein 
gamma and zeta function regulation is employed.  The key reason for the surprising 
accuracy of the ideal boson and ideal fermion cases resides in the asymptotic freedom 
or equivalently the Feynman ``parton'' structure of the ultra-high energy tails of spectral 
functions. 
\end{abstract}


\maketitle

\section{Introduction \label{intro}}

As in our preliminary work\cite{widom:2014}, we seek to understand the nature of the 
power law exponents \begin{math} \{\alpha \}  \end{math} which are employed to describe 
the energy distributions observed in the cosmic rays continually bombarding our planet 
and coming from astrophysical sources\cite{Gaisser:1990,Stanev:2004,PDG:2014}. From the 
quantum field theory viewpoint we regard the cosmic rays as standard model hadrons 
evaporating  from sources and moving away from such sources as a gaseous blowing 
wind. Such a {\em solar} wind exists issuing from the center of our own planetary system. 
These evaporating winds no doubt also blow away from other astrophysical objects such 
as neutron stars. 

The starting point for defining cosmic ray power law exponents was purely experimental. It is 
known\cite{Gaisser:2002} that the energy distribution law of cosmic ray nuclei in the 
energy range \begin{math}  5 {\rm \ GeV} < E < 100 {\rm \ TeV} \end{math} via the 
differential flux per unit time per unit area per steradian per unit energy obeys
\begin{equation}
\frac{d^4 N}{dt dA d\Omega dE} \approx 
\frac{1.8\ {\rm nucleons}}{\rm sec\ cm^2\ sr\ GeV}
\left(\frac{\rm 1\ GeV}{E}\right)^\alpha 
\label{intro1}
\end{equation}
wherein the experimental power law exponent \begin{math} \alpha \approx 2.7  \end{math}.
At the ``knee'' of the distribution, i.e. at energy 
\begin{math} E\sim 1\ {\rm PeV} \end{math}, there is a shift in the power law exponent to 
the value \begin{math} \alpha \approx 3.1  \end{math}. 
In \cite{widom:2014}, we had computed theoretically the ideal Bose index. Here we also
compute the ideal Fermi statistical index so that they read together as: 
\begin{equation}
\alpha_{\rm Bose}=2.701178\ \ \ {\rm and}
\ \ \ \alpha_{\rm Fermi}=3.151374\ . 
\label{intro2}
\end{equation}
It would be well within experimental error to regard the knee as a crossover between 
statistics which in concrete physical evaporation terms merely means a crossover 
in the composition of cosmic ray emission winds blowing away from astrophysical 
sources. The critical values in Eq.(\ref{intro2}) are {\em ideal} in the sense that 
the particles are ultra-relativistic \begin{math} E \approx c|{\bf p}|  \end{math} and 
noninteracting. One might ponder why a non-interacting theory is so close to 
experimental reality. The answer resides in the asymptotic freedom 
in the form of Feynman parton structure\cite{Feynman:1969} of the ultra-high 
energy tails of spectral functions. 

To describe cosmic ray sources in terms of thermal quantum field theoretical models, 
it is of some convenience to employ gamma and zeta function regulators whose 
definitions are reviewed in Sec.\ref{gzfr} wherein the ideal power law exponents are derived.
That interactions apparently have little effect on the power law exponents would seem to imply 
that the quantum spectral functions are of the Feynman form\cite{{Feynman:1969}} with 
Bose and Fermi operators being composites of quark operators. In the concluding 
Sec.\ref{conc} these points are qualitatively discussed. 

\section{Gamma and Zeta Regulators \label{gzfr}}

\subsection{Mathematical Details \label{mc}}

The mathematics of gamma and zeta regulators resides in the properties of classical
special functions\cite{nist:2010}. Starting with the statistical index 
\begin{equation}
\eta=1\ \ {\rm Bose},\ \ \eta=0\ \ {\rm Boltzmann}\ \ {\rm and}\ \ \eta=-1\ \ {\rm Fermi},   
\label{gzfr1}
\end{equation}
the general  zeta function regulators are defined as 
\begin{equation}
Z(s,\eta )=\int_0^\infty \frac{x^s}{e^x-\eta }\left[\frac{dx}{x}\right] 
\ \ {\rm for} \ \ \Re e\  s > 1
\label{gzfr2}
\end{equation}
and by analytic continuation in {\it s} elsewhere. The Boltzmann regulator is the 
Euler gamma function
\begin{equation}
Z(s,0)=\int_0^\infty x^s e^{-x}\left[\frac{dx}{x}\right]=\Gamma (s).
\label{gzfr3}
\end{equation}
The Bose regulator is determined by the Riemann zeta function defined by
\begin{equation}
\zeta (s)=\sum_{n=1}^\infty \frac{1}{n^s}
\end{equation}
or its analytic continuation in {\it s} where the sum does not converge
via 
\begin{equation}
Z(s,1)=\int_0^\infty x^s 
\sum_{n=1}^\infty e^{-ns}\left[\frac{dx}{x}\right]=\Gamma (s)\zeta (s).
\label{gzfr4}
\end{equation}
The Fermi regulator 
\begin{eqnarray}
Z(s,-1)=\int_0^\infty x^s 
\sum_{n=1}^\infty (-1)^{n-1}  e^{-ns}\left[\frac{dx}{x}\right], 
\nonumber \\ 
Z(s,-1)=\Gamma (s)\sum_{n=1}^\infty \frac{(-1)^{n-1}}{n^s}\ ,
\nonumber \\ 
Z(s,-1)=\left[1-\frac{2}{2^s}\right]\Gamma (s)\zeta(s). 
\label{gzfr5}
\end{eqnarray}
Note that the Bose and Fermi regulators are rigorously yet simply 
related by 
\begin{equation}
Z(s,-1)=\left[1-\frac{2}{2^s}\right]Z(s,1).
\label{gzfr5}
\end{equation}

\subsection{Ideal power law exponents \label{idi}}

The density of states per unit energy per unit volume for ultra-relativistic 
particles is proportional to the square of the energy. The mean energy 
per particle in an ideal gas is thereby 
\begin{equation}
E_\eta = \frac{\int_0^\infty \frac{\epsilon^3 d\epsilon }{e^{\epsilon/k_BT}-\eta}}
{\int_0^\infty \frac{\epsilon^2 d\epsilon }{e^{\epsilon/k_BT}-\eta}}=
\alpha_\eta k_BT,
\label{gzfr6}
\end{equation}
wherein the regulators determining the mean energies are 
\begin{equation}
\alpha_\eta =\left[\frac{Z(4,\eta )}{Z(3,\eta )}\right].
\label{gzfr7}
\end{equation}
In detail, 
\begin{eqnarray} 
\alpha_0 =\frac{\Gamma(4)}{\Gamma(3)}=\frac{3!}{2!}=3,
\nonumber \\ 
\alpha_1=\frac{\Gamma(4)\zeta(4)}{\Gamma(3)\zeta(3)}
\approx 2.701178, 
\nonumber \\ 
\alpha_{-1}=\left[\frac{1-(1/8)}{1-(1/4)} \right]\alpha_1=\frac{7\alpha_1}{6}
\approx 3.151374,
\label{gzfr8}
\end{eqnarray}
as in Eq.(\ref{intro2}).

To establish \begin{math}  \alpha  \end{math} as a power law exponent when the 
energy \begin{math}  E=\alpha k_BT  \end{math}, one must compute 
the entropy as 
\begin{equation}
E=\alpha k_BT=\alpha k_B \frac{dE}{dS} \ \ \Rightarrow 
\ \ S=\alpha k_B\ln \left(\frac{E}{E_0}\right)
\label{gzfr9}
\end{equation}
and employ the heat of vaporization to compute the evaporation 
energy spectrum
\begin{equation}
e^{-S/k_B}=\left(\frac{E_0}{E}\right)^\alpha 
\label{gzfr10}
\end{equation}
as in Eq.(\ref{intro1}).
 
\section{Conlusions \label{conc}}

 A more detailed interacting quantum field theoretical calculation of 
\begin{math} \{\alpha \}  \end{math} power law exponents involves the 
construction of single particle spectral functions in the context 
of thermal quantum field theory. While we have here computed
the \begin{math} \{\alpha \}  \end{math} indices for the free Fermi 
and free Bose field theories, the results are already in quite satisfactory 
agreement with experimental cosmic ray power law exponents. The reason 
for this remarkable agreement would appear to be due to a Feynman 
``parton" structure\cite{Feynman:1969} for the high energy asymptotic 
tails of the single particle spectral functions. In this case that structure 
would be described by free non-interacting particles. Following 
Feynman's physical reasoning and employing the dispersion relations 
in a finite temperature many body quantum field theory context, we are 
presently computing the renormalized energy dependent power law exponent 
\begin{math}  \alpha (E)  \end{math} for interacting theories.

\section{Acknowledgements}

J.S. thanks the National Science Foundation for its support via NSF grant PHY-1205845.

\end{document}